\documentclass[prd,aps,preprintnumbers,floats,floatfix,superscriptaddress,preprintnumbers,showpacs,eqsecnum,nofootinbib,notitlepage,twocolumn]{revtex4-1}
\usepackage[utf8]{inputenc}
\usepackage{latexsym,array,theorem,mathrsfs,bm,float}
\usepackage{psfrag}
\usepackage{amsfonts,amsmath,amssymb,latexsym,array,afterpage,
theorem,mathrsfs,bm,float,epsfig,color,graphicx,tabularx,here,multirow,hyperref}

\newcommand{\nn}{\nonumber \\}

\newcommand{\Xcal}{ \mathcal{X} }

\newcommand{\Mcal}{ \mathcal{M} }

\newcommand{\inta}{\int_{ \Mcal_3} \!\!\!\!\!}
\newcommand{\intb}{\int_{\mathcal{M}_4} \!\!\!\!\!}

\newcommand{\Mpl}{M_{\rm pl}}

\newcommand{\one}{{\bm 1}}

\begin{document}
\baselineskip=12pt

\preprint{YITP-20-119}
\title{Non-linearly ghost-free higher curvature gravity}
\author{Katsuki Aoki}
\email{katsuki.aoki@yukawa.kyoto-u.ac.jp}
\affiliation{Center for Gravitational Physics, Yukawa Institute for Theoretical Physics, Kyoto University, 606-8502, Kyoto, Japan}

\date{\today}

 \begin{abstract}
We find unitary and local theories of higher curvature gravity in the vielbein formalism, known as the Poincar\'{e} gauge theory by utilizing the equivalence to the ghost-free massive bigravity. We especially focus on three and four dimensions but extensions into a higher dimensional spacetime are straightforward. In three dimensions, a quadratic gravity $\mathcal{L}=R+T^2+R^2$, where $R$ is the curvature and $T$ is the torsion with indices omitted, is shown to be equivalent to zwei-dreibein gravity and free from the ghost at fully non-linear orders. In a special limit, new massive gravity is recovered. When the model is applied to the AdS/CFT correspondence, unitarity both in the bulk theory and in the boundary theory implies that the torsion must not vanish. On the other hand, in four dimensions, the absence of ghost at non-linear orders requires an infinite number of higher curvature terms, and these terms can be given by a schematic form $R(1+R/\alpha m^2)^{-1}R$ where $m$ is the mass of the massive spin-2 mode originating from the higher curvature terms and $\alpha$ is an additional parameter that determines the amplitude of the torsion. We also provide another four-dimensional ghost-free higher curvature theory that contains a massive spin-0 mode as well as the massive spin-2 mode.
 \end{abstract}

\maketitle
\section{Introduction}
Giving a mass to the graviton is an idea from the pioneering study by Fierz and Pauli~\cite{Fierz:1939ix} and has been extensively discussed since then  (see~\cite{deRham:2014zqa,Schmidt-May:2015vnx} for reviews). Although it was believed that appearance of a pathological mode, called the Boulware-Deser (BD) ghost~\cite{Boulware:1973my}, is inevitable in theories of massive spin-2 field, the ghost-free non-linear extension of the Fierz-Pauli theory was discovered by de Rham et al.~in 2010~\cite{deRham:2010ik,deRham:2010kj}. This ghost-free theory, often dubbed the de Rham-Gabadadze-Tolley (dRGT) theory, is the theory of a single massive spin-2 field and was extended to the bigravity~\cite{Hassan:2011zd} that includes the massless spin-2 field as well, and to the multi-gravity that has multiple massive spin-2 fields as well as the massless one~\cite{Hinterbichler:2012cn}. The dRGT theory and its extensions provide a well-defined framework of modification of general relativity (GR) and many phenomenological aspects have been investigated. If a future observation reveals the existence of the massive spin-2 field(s), what can we learn about the nature of gravity? In the case of particle physics, the discovery of the massive gauge bosons was a road to revealing the standard model of particle physics. To answer this question, we should investigate the underlying physics of the theories of massive gravity. The simplest possibility would be a higher dimensional scenario. The paper~\cite{deRham:2013awa} indeed demonstrates that the ghost-free theories can be obtained as a Kaluza-Klein compactification of the higher dimensional GR. In this case, multiple massive spin-2 fields are naturally expected, corresponding to the Kaluza-Klein states of the graviton.

Another previous idea was to interpret gravity as a gauge force initiated by Utiyama~\cite{Utiyama:1956sy}, Kibble~\cite{Kibble:1961ba}, and Sciama~\cite{sciama1962analogy} (see~\cite{Blagojevic:2002du,Blagojevic:2013xpa} for reviews). In particle physics, force and symmetry are closely related. The idea of the gauge principle explains elegantly the existence of gauge fields and interactions between the gauge fields and matter fields. One may apply the same idea to gravity. Since a particle is specified by energy, momentum and angular momentum (spin), namely currents associated with the Poincar\'{e} group, it would be natural to suppose that gravity is a Poincar\'{e} gauge theory (PGT). The corresponding gauge fields are the vielbein $e^a_{\mu}dx^{\mu}$ and the spin connection $\omega^{ab}{}_{\mu}dx^{\mu}$ which are regarded as the gauge fields associated with the translation part and the rotation part of the Poincar\'{e} group, respectively. Let us then consider the dynamics of the gauge fields. The field strengths are the curvature and the torsion. An important observation is that PGT can be interpreted as a theory in a Higgs phase~\cite{Percacci:1990wy,Percacci:2009ij}. When one considers a general Lagrangian consisting of the curvature and the torsion, and studies perturbations around the flat background, $e^a_{\mu}=\delta^a_{\mu},\omega^{ab}{}_{\mu}=0$, one finds that the particles associated with the spin connection are massive. Hence, in a low energy limit, the spin connection can be integrated out and the vielbein is the only dynamical field. Thanks to the local Lorentz invariance, we can use the spacetime metric as the independent variable of gravity instead of the vielbein. The leading piece of the Lagrangian in the low energy limit is the Einstein-Hilbert action. As a result, GR arises naturally as a low energy effective field theory (EFT) of PGT.  However, similarly to the weak force, a new force carried by the massive spin connection should appear at short distances.

There have been many attempts to identify a viable model of PGT involving a dynamical spin connection. When the Lagrangian of PGT is supposed to be algebraic in the field strengths, the existence of first-class constraints, namely the symmetries of PGT, concludes that there exist 18 degrees of freedom (dofs) in addition to the massless graviton in four dimensions~\cite{Blagojevic:1983zz,Nikolic:1984xi}. These 18 dofs are classified into those of massive spin-$2^{\pm},1^{\pm},0^{\pm}$ particle species where the number and $\pm$ of $J^P~(J=0,1,2,~P=\pm)$ denote the spin and the parity, respectively. Since the equations of motion are second order, there is no Ostrogradsky ghost. However, all of these particle species cannot be physical, simultaneously~\cite{Sezgin:1979zf,Sezgin:1981xs} (see~\cite{Karananas:2014pxa,Blagojevic:2018dpz,Lin:2018awc,Aoki:2019snr} for recent discussions). This can be understood by the fact that the Poincar\'{e} group is non-compact, meaning that the norm is not positive definite. There can be a negative (ghostly) mode, in general. We should fine-tune the Lagrangian so that the masses of the ghost modes are sufficiently heavy; we can then discuss dynamics of (physical modes of) the spin connection.
Even though we need to accept the fine-tuning, the idea of PGT is natural and worth considering. In particular, if the existence of the dynamical spin connection is revealed by an experiment, it can be described as the discovery of ``massive gauge bosons'' in the gravity sector.

In the present paper, we connect two ideas, massive gravity and PGT, and show that a particular class of PGT is equivalent to the ghost-free bigravity in a vacuum. Similar discussions were already made in~\cite{Paulos:2012xe,Gording:2018not}; however, an important difference is that we show the equivalence without violating either unitarity or locality. From the perspective of massive gravity, we can interpret the massive spin-2 field as a ``massive gauge boson'' arising from a higher curvature correction to GR. It is well-known that higher curvature corrections to GR give rise to a massive spin-2 mode but this mode is a ghost mode according to the Ostrogradsky theorem and cannot be thought of as a physical state of the theory. However, the torsion enables us to find a ghost-free higher curvature theory as we will see. On the other hand, the equivalence provides a counterexample to a speculated conclusion of PGT: only good propagating modes of the spin connection are the spin-0 modes. Although one can find a ghost-free PGT at the level of linear perturbations about the flat background, the non-linear interactions may drastically change the structure of the theory. According to~\cite{Yo:1999ex,Yo:2001sy}, it seems that the non-linearly ghost-free PGT can only contain the spin-0 modes in addition to the massless graviton (see also~\cite{Jimenez:2019qjc}). However, there is a non-linearly ghost-free PGT involving a massive spin-2 mode when an infinite number of appropriate higher curvature terms are added. The ghost-free higher curvature terms are determined by a finite number of coupling constants because of the restriction of the dRGT mass terms.

The rest of the present paper is organized as follows. We first summarise notation and definitions in Section~\ref{sec_notation}. Section~\ref{sec_3d} is devoted to studying a three-dimensional quadratic gravity. The consideration of the three-dimensional gravity not only provides a simplified relation between the higher curvature gravity and the bigravity but also provides an interesting insight related to the AdS/CFT correspondence. We then study four-dimensional higher curvature theories in Section~\ref{sec_4d} and find that the fully ghost-free higher curvature gravity is obtained by adding an infinite number of appropriate higher curvature terms. In Section~\ref{sec_add_scalar}, we consider an extension of the argument of Section~\ref{sec_4d}. We conclude in Section~\ref{summary} with a summary and discussions.

\section{Notation}
\label{sec_notation}
Throughout the present paper, Greek indices $\mu,\nu,\cdots$ are used to denote spacetime indices whereas Latin indices $a,b,\cdots$ are used to represent the Lorentz indices. The signature of the metric is $(-,+,\cdots,+)$ and $\eta_{ab}$ is the Minkowski metric. Only in this section, we consider $D$-dimensional spacetime and we will focus on three and four dimensions in the latter sections.

The basic variables of the PGTs are the vielbein $e^a=e^a_{\mu}dx^{\mu}$ and the spin connection $\omega^{ab}=\omega^{ab}{}_{\mu}dx^{\mu}$ which are regarded as gauge fields associated with the translation and the Lorentz transformations, respectively. Since we are interested in the $SO(1,D-1)$ connection, the spin connection is assumed to have anti-symmetric indices, $\omega^{ab}=\omega^{[ab]}$. The associated field strengths are the torsion 2-form $T^a=\frac{1}{2}T^{a}{}_{\mu\nu}dx^{\mu}\wedge dx^{\nu}$ and the curvature 2-form $R^{ab}=\frac{1}{2}R^{ab}{}_{\mu\nu}dx^{\mu}\wedge dx^{\nu}$ defined by
\begin{align}
T^a&:=De^a =de^a+\omega^a{}_b \wedge e^b
\,, \\
R^{ab}&:=d\omega^{ab}+\omega^a{}_{c}\wedge \omega^{cb}
\,,
\end{align}
where $d$ and $D$ are the exterior derivative and the exterior covariant derivative, respectively. The Levi-Civita tensor is denoted by $\epsilon_{a_1 a_2 \cdots a_D}$ with $\epsilon_{012\cdots }=1$. In component expressions, the spacetime indices are converted to the Lorentz indices vice versa with the help of the vielbein and its inverse.

The Ricci tensor and the Ricci scalar are defined by
\begin{align}
R^a{}_{\mu}:=R^{ba}{}_{\nu\mu}e_b^{\nu} \,, \quad R:=R^{a}{}_{\mu}e^{\mu}_a
\,,
\end{align}
where the tensor $R_{\mu\nu}=e^a_{\mu}R_{a\nu}$ is not symmetric in its indices contrary to the Riemannian space. We then introduce the Schouten 1-form
\begin{align}
S^a=S^a{}_{\mu}dx^{\mu}:=\left(R^a{}_{\mu}-\frac{1}{2(D-1)}Re^a{}_{\mu} \right)dx^{\mu}
\,,
\end{align}
and the contortion 1-form
\begin{align}
K^{ab}=K^{ab}{}_{\mu}dx^{\mu}&:=\frac{1}{2}(T_{\mu}{}^{ab}-T^{b}{}_{\mu}{}^{a}-T^{ab}{}_{\mu})dx^{\mu}
\,. 
\end{align}
The contortion tensor represents the difference between the general connection and the torsionless connection,
\begin{align}
K^{ab}{}_{\mu}=\omega^{ab}{}_{\mu}-\omega^{ab}{}_{\mu}(e)
\label{contortion}
\end{align}
where $\omega^{ab}{}_{\mu}(e)$ is the torsionless spin connection which is determined by the vielbein. Similarly, we denote the curvature tensors as $R^{ab}{}_{\mu\nu}(e),R^a{}_{\mu}(e),R(e)$ when they are determined by the torsionless connection  $\omega^{ab}{}_{\mu}(e)$. Later, we will introduce another vielbein $f^a_{\mu}$. The curvature 2-form with respect to $f^a_{\mu}$ is denoted by $R^{ab}(f)$.

\section{Three dimensions}
\label{sec_3d}
As mentioned, in the general Lagrangian, there must be ghostly dofs because the positive definiteness is not guaranteed. We need a fine-tuning in order that the ghost modes do not propagate at least in a low energy regime. If we demand that the Lagrangian is at most quadratic in derivatives, the Lagrangian is schematically given by the form $\mathcal{L}=R+T^2+R^2$, called quadratic PGTs. At the level of linear perturbations around the flat background, discussions about the particle contents and the ghost-free conditions of the three-dimensional quadratic PGTs can be found in \cite{Hernaski:2009wp,HelayelNeto:2010jn}. In this section, among general possibilities of PGTs, we focus on a specific model of the quadratic PGT which only has a massive spin-2 dof in the linear perturbations.

Let us consider the following Lagrangian of a higher curvature gravity in $D=3$ dimensions:
\begin{align}
S&=M_3  \inta L
\nn
L&= \epsilon_{abc} \left[ \sigma e^a \wedge R^{bc} - \frac{\alpha}{2} T^a \wedge K^{bc} - \frac{1}{2M_*^2} S^a \wedge R^{bc} \right]
,
\label{TNMG}
\end{align}
where $M_3$ is the three-dimensional Planck mass, $M_*$ is a mass parameter, $\sigma=\pm 1$, and $\alpha$ is a dimensionless coupling constant. For simplicity, we ignore the cosmological constant but the inclusion of the cosmological constant is straightforward. The ``wrong'' sign $\sigma=-1$ is allowed because the massless spin-2 field has no degrees of freedom in three dimensions. The Lagrangian consists of particular torsion square and curvature square terms, $T\wedge K$ and $S\wedge R$, in addition to the Einstein-Hilbert term $e\wedge R$. These particular combinations ensure that the Lagrangian \eqref{TNMG} only has the massive spin-2 mode and is free from the ghosts at the linear level of perturbations (see e.g.~\cite{Sezgin:1979zf,Sezgin:1981xs,Hernaski:2009wp,HelayelNeto:2010jn}). A remarkable feature is that even at fully non-linear orders these combinations are free from the ghostly dofs as we will see.

The (dualized) equations of motion are given by
\begin{align}
\sigma R^{ab}-\alpha(DK^{ab}-K^{a}{}_c \wedge K^{cb}) +\frac{1}{M_*^2}S^{[a}\wedge S^{b]}&=0
\,, \label{eq_drei} \\
(\alpha-\sigma) T^a + \frac{1}{M_*^2}DS^a&=0
\,. \label{eq_spin}
\end{align}

When we ignore the coupling $T\wedge K$ and the torsion, the model is knowns as new massive gravity (NMG)~\cite{Bergshoeff:2009hq}. Note, however, that the $\alpha \to 0$ limit of \eqref{TNMG} does not lead to NMG; rather, NMG is obtained as the $|\alpha| \to \infty$ limit of \eqref{TNMG}. As shown in \eqref{contortion}, the general connection can be divided into the the torsionless part $\omega^{ab}(e)$ and the torsion part $K^{ab}$. We can thus use the torsion and the dreibein as independent variables instead of the dreibein and the spin connection. In this case, the second term $T\wedge K$ can be regarded as a ``mass'' term of the torsion. Therefore, the limit $|\alpha| \to \infty$ is an infinitely heavy limit of the torsion. The torsion can be integrated out and a local action can be obtained by expanding it in terms of $\alpha^{-1}$. The perturbative solution of \eqref{eq_spin} under $|\alpha| \to \infty$ is
\begin{align}
T^a=-\frac{1}{\alpha M_*^2 }D(e) S^a(e)+\mathcal{O}(\alpha^{-2})
\end{align}
where $D(e)$ is the exterior covariant derivative defined by the torsionless connection $\omega^{ab}(e)$.
As a result, the action becomes
\begin{align}
S&= M_3  \inta \epsilon_{abc} \left[ \sigma e^a \wedge R^{bc}(e) - \frac{1}{2M_*^2} S^a(e) \wedge R^{bc}(e) \right] 
\nn
&+\mathcal{O}(\alpha^{-1})
\nn
&=M_3 \inta d^3x |e| 
\nn
&\times \left[ \sigma R(e)+\frac{1}{M_*^2}\left(R_{\mu\nu}(e)R^{\mu\nu}(e)-\frac{8}{3}R^2(e) \right) \right]+\mathcal{O}(\alpha^{-1})
\,,
\label{NMG}
\end{align}
and then the original NMG is obtained under the limit $|\alpha| \to \infty$. In NMG, we should assume $\sigma=-1$ as required by unitarity. The massive spin-2 mode is then a non-ghost particle. Hereinafter we shall call the model \eqref{TNMG} torsional new massive gravity (TNMG) whereas the model \eqref{NMG} is simply called NMG.

It is known that NMG is obtained as a scaling limit of the three-dimensional ghost-free bigravity theory, dubbed zwei-dreibein gravity~\cite{Paulos:2012xe,Bergshoeff:2013xma}. A point of the present paper is that the equivalence between TNMG and the ghost-free bigravity is shown without the use of any scaling limit. As mentioned, NMG is obtained as a particular limit of TNMG and this limit is indeed what the papers \cite{Paulos:2012xe,Bergshoeff:2013xma} considered.

The idea is basically the same as what is done in the analysis of NMG~\cite{Bergshoeff:2009hq,Paulos:2012xe,Bergshoeff:2013xma} or of $f(R)$ theories but we need to elaborate the way of inclusion of auxiliary variables. Let us introduce two auxiliary 1-forms $\xi^a$ and $\chi^{ab}=\chi^{[ab]}$, and rewrite the action \eqref{TNMG} as an equivalent form
\begin{align}
S_{\rm eq}&=M_3 \inta L_{\rm eq}
\,, \nn
L_{\rm eq}&=\epsilon_{abc}\big[\sigma e^a \wedge R^{bc}+\alpha (T^a\wedge \chi^{bc}+e^a \wedge \chi^b{}_d \wedge \chi^{dc})
\nn
&\qquad \quad
+(\sigma-\alpha) \xi^a \wedge R^{bc}+ (\sigma-\alpha)^2 M_*^2 e^a\wedge \xi^b \wedge \xi^c
\big]
. \label{eqTNMG}
\end{align}
Since the variables $\xi^a$ and $\chi^{ab}$ are auxiliary ones, the equations of motion of them yield
\begin{align}
\xi^a=\frac{1}{(\alpha-\sigma) M_*^2}S^a\,,\quad \chi^{ab}=-K^{ab}
\,, \label{aux_sol}
\end{align}
and then the original action \eqref{TNMG} is recovered by substituting the solutions into \eqref{eqTNMG}. Instead, we perform integration by parts to find
\begin{align}
L_{\rm eq}=\epsilon_{abc}\big[
&\sigma e^a \wedge R^{bc}+\alpha e^a \wedge (D\chi^{bc}+\chi^b{}_d \wedge \chi^{dc})
\nn
&+(\sigma-\alpha) \xi^a \wedge R^{bc}+ (\sigma-\alpha)^2 M_*^2 e^a\wedge \xi^b \wedge \xi^c
\big]
, \label{eqTNMG2}
\end{align}
where we have used $T^a=De^a$.
Then, we introduce a new dreibein and a new spin connection via
\begin{align}
f^a=e^a+ \xi^a
\,, \quad 
\Omega^{ab}=\omega^{ab}+\chi^{ab}
\end{align}
and define the curvature 2-form associated with the new connection,
\begin{align}
F^{ab}:=d\Omega^{ab}+\Omega^{a}{}_c \wedge \Omega^{bc}
\,,
\end{align}
which yields the relation
\begin{align}
D\chi^{ab}+\chi^{a}{}_c \wedge \chi^{cb}=F^{ab}-R^{ab}
\,. \label{dif_cur}
\end{align}
It is straightforward to show that the action \eqref{eqTNMG2} is written as the form of the ghost-free bigravity, also called zwei-dreibein gravity,
\begin{align}
L_{\rm eq}= \epsilon_{abc}\big[ &\alpha e^a \wedge F^{bc}+ (\sigma-\alpha) f^a \wedge R^{bc}
\nn
& + (\sigma-\alpha)^2 M_*^2 e^a\wedge (f^b-e^b) \wedge (f^c-e^c) \big]
\,.
\label{bigravity_3d}
\end{align}

Let us confirm the consistency with the original formulation \eqref{TNMG}. The equations of motion of \eqref{bigravity_3d} are
\begin{align}
\alpha F^{ab}-(\sigma -\alpha)^2M_*^2 (2e^{[a} \wedge \xi^{b]}-\xi^{[a}\wedge \xi^{b]})&=0
\,, \label{g_Ein_eq} \\
R^{ab}+2(\sigma-\alpha)M_*^2 e^{[a}\wedge \xi^{b]}&=0 \label{f_Ein_eq}
\,,
\end{align}
and
\begin{align}
de^a+\Omega^a{}_b \wedge e^b&=0 \,, \label{g_torsionless} \\
df^{a}+\omega^a{}_b\wedge f^b&=0\,, \label{f_torsionless}
\end{align}
where $\xi^a=f^a-e^a$. The equations \eqref{f_Ein_eq} and \eqref{g_torsionless} lead to the equations \eqref{aux_sol}. Substituting \eqref{aux_sol} into the equations \eqref{g_Ein_eq} and \eqref{f_torsionless}, we obtain the original equations of motion \eqref{eq_drei} and \eqref{eq_spin}. Therefore, two systems \eqref{TNMG} and \eqref{bigravity_3d} are indeed equivalent.

\begin{table}
\caption{Summary of the variables} \label{table_var}
\begin{tabular}{cccc} \hline
                     & variables                 & curvature   & torsion \\ \hline\hline
physical variables & $(e^a,\omega^{ab})$   & $R^{ab}$    & $T^a \neq 0$ \\
$g$-variables      & $(e^a,\Omega^{ab})$   & $F^{ab}$   & $de^a+\Omega^a{}_b \wedge e^b=0 $ \\
$f$-variables     & $(f^a,\omega^{ab})$    & $R^{ab}$    & $df^a+\omega^a{}_b \wedge f^b =0 $ \\ \hline
\end{tabular}
\end{table}

Using \eqref{g_torsionless} and \eqref{f_torsionless}, we can eliminate the spin connections from the action as with the Einstein-Cartan-Sciama-Kibble theory. We then obtain the torsionless bigravity
\begin{align}
L_{\rm eq}=\epsilon_{abc}\big[ &\alpha e^a \wedge R^{bc}(e)+ (\sigma-\alpha) f^a \wedge R^{bc}(f) 
\nn
&+ (\sigma-\alpha)^2 M_*^2 e^a\wedge (f^b-e^b) \wedge (f^c-e^c) \big]
. \label{bigravity_3d2}
\end{align}
It would be worth emphasising that in the original action \eqref{TNMG}, the Einstein-Hilbert term is composed of the pair $(e^a,\omega^{ab})$ while a couple of Einstein-Hilbert action in \eqref{bigravity_3d} consist of the different pairs $(e^a,\Omega^{ab})$ and $(f^a,\omega^{ab})$. We summarise these three pairs of the variables, which we shall call physical, $g$-, and $f$-variables, in Table \ref{table_var}.
The ``torsionless'' conditions \eqref{g_torsionless} and \eqref{f_torsionless}
do not yield the torsionless condition $T^a=de^a+\omega^a{}_b \wedge e^b=0$ in the sense of the  higher curvature theory \eqref{TNMG}. The torsionless bigravity \eqref{bigravity_3d2} is equivalent to the torsionfull higher curvature theory \eqref{TNMG}.
The symmetric dreibein condition $e^a \wedge f_a=0$ is obtained as a constraint of \eqref{bigravity_3d2} under which the bigravity in the dreibein formalism is reduced to that in the metric formalism~\cite{Nibbelink:2006sz,Hinterbichler:2012cn,Deffayet:2012zc}.\footnote{In general, there can exist other branches of the constraint~\cite{Hinterbichler:2012cn,Banados:2013fda}. The equivalence between the bigravity in the vielbein formalism and that in the metric formalism holds only when one chooses the branch $e^a \wedge f_a=0$ of the constraint. Throughout the present paper, we only consider the branch $e^a \wedge f_a=0$ to guarantee the absence of the BD ghost. } The theory is then defined by two ``metrics'' $g_{\mu\nu}=\eta_{ab}e^a_{\mu}e^b_{\nu}$ and $f_{\mu\nu}=\eta_{ab}f^a_{\mu}f^b_{\nu}$ where the $g$-metric coincides with the physical one of the original theory \eqref{TNMG} while the $f$-metric appears as a result of the inclusion of the auxiliary variables. The symmetric dreibein condition $e^a \wedge f_a=0$ reads that the physical Ricci tensor, which is defined by the pair $(e^a,\omega^{ab})$, is symmetric
\begin{align}
R_{[\mu\nu]}=0
\,,
\end{align}
although the torsion does not vanish.

NMG \eqref{NMG} is obtained as the $|\alpha| \to \infty$ limit of \eqref{TNMG}. Since we have proved TNMG is equivalent to the ghost-free bigravity, we can also conclude that NMG is obtained as a limit of the bigravity which was already pointed out in the paper~\cite{Paulos:2012xe} (see also \cite{Bergshoeff:2013xma}). Let us consider the scaling limit investigated in~\cite{Paulos:2012xe}. To follow the paper~\cite{Paulos:2012xe}, we first rewrite the action \eqref{bigravity_3d2} as
\begin{align}
S_{\rm eq}= \inta \epsilon_{abc}\big[ 
&M_g e^a \wedge R^{bc}(e)+ M_f f^a \wedge R^{bc}(f) 
\nn
&+  M_{\rm eff} m^2 e^a\wedge (f^b-e^b) \wedge (f^c-e^c) \big]
,
\end{align}
where
\begin{align}
M_g&:=\alpha M_3\,,\quad M_f:=(\sigma-\alpha)M_3\,, \nn
M_{\rm eff}&:=(M_g^{-1}+M_f^{-1})^{-1}=\frac{\alpha(\sigma-\alpha)}{\sigma}M_3
\,,
\end{align}
and
\begin{align}
m^2:=\frac{\sigma(\sigma-\alpha)}{\alpha}M_*^2
\end{align}
is the mass of the spin-2 mode about the flat background.
The paper~\cite{Bergshoeff:2013xma} considered the limit
\begin{align}
M_f \to + \infty
\end{align}
keeping $-(M_g+M_f)>0$ and $m$ fixed. Clearly, this limit is equivalent to the $\alpha \to -\infty$ limit with $\sigma=-1$ keeping $M_3,M_*$ fixed. We thus reproduce the result of~\cite{Bergshoeff:2013xma} in a different way. One can also consider the limit $M_f \to -\infty$ keeping $-(M_g+M_f)>0$ and $m$ fixed which corresponds to the limit $\alpha \to +\infty$. In the NMG limit $|\alpha|\to \infty$, one of the ``Planck masses'', $M_g$ or $M_f$, has to be negative.

The higher curvature term may be interpreted as a correction of quantum gravity which motivates us to think of the AdS/CFT correspondence in which a $D$-dimensional quantum gravity in an asymptotically anti-de Sitter (AdS) spacetime is conjectured to be equivalent to a $(D-1)$-dimensional conformal field theory (CFT).
To discuss the implication for the AdS/CFT correspondence, we add the cosmological constant term 
\begin{align}
-M_3 \inta ~ \epsilon_{abc} \frac{1}{3}\lambda M_*^2 e^a \wedge e^b \wedge e^c
\label{cc_term}
\end{align}
to the action \eqref{TNMG} for a while, where $\lambda$ is dimensionless. The same term is then added to \eqref{bigravity_3d} after changing the variables.
In three dimensions, the ghostly massless spin-2 field does not violate unitarity. Therefore, the limit $|\alpha| \to \infty$ namely NMG provides a well-defined lower-dimensional gravity as long as the massive spin-2 field is not a ghost. However, in NMG, there is a conflict between unitarity in the bulk theory and unitarity in the dual CFT as required by the positive central charge~\cite{Bergshoeff:2009aq,Liu:2009bk,Liu:2009kc}. 
The paper~\cite{Bergshoeff:2013xma} then proposed zwei-dreibein gravity to resolve this problem. Nonetheless, in the generic parameter space, zwei-dreibein gravity cannot be recast in the form of a higher curvature gravity in the metric formalism unless introducing infinite derivatives (see e.g.~\cite{Gording:2018not}).

We now know the equivalence between zwei-dreibein gravity and TNMG. We can understand, in the language of the higher curvature theory, what was the missing ingredient of NMG to preserve unitarity in both the bulk theory and the dual CFT.  
After adding \eqref{cc_term} and substituting the ansatz of maximally symmetric spacetime into the equations of motion, we obtain
\begin{align}
R^{a}{}_{\mu}-\frac{1}{2}Re^a_{\mu}+\Lambda e^a_{\mu}=0
\,, \quad
T^a{}_{\mu\nu}=0
\end{align}
where
\begin{align}
\Lambda&=-2M_*^2(\sigma \pm \sqrt{1+\lambda} )
\,.
\end{align}
The constant $\Lambda$ determines the cosmological constant of the maximally symmetric solution. In this solution, the $f$-dreibein is given by $f^a=\gamma e^a$ where 
\begin{align}
\gamma=1+\frac{\Lambda}{2M_*^2(\alpha-\sigma)}
\,.
\end{align}
Following the paper~\cite{Bergshoeff:2013xma}, the ghost-free and tachyon-free (bulk unitarity) conditions around the AdS background require
\begin{align}
\frac{\alpha(\sigma-\Lambda/2M_*^2)}{\sigma-\alpha-\Lambda/2M_*^2}&>0 \,,  \label{no_ghost}\\
\mathcal{M}^2:=m^2+\frac{1}{2}\Lambda\left(1-\frac{\Lambda}{2\alpha M_*^2} \right)&>0
\,,
\end{align}
while the positive central charge (boundary unitarity) requires
\begin{align}
c:=24 \pi \ell M_3 \left(\sigma -\frac{\Lambda}{2M_*^2} \right)>0
\,, \label{central_charge}
\end{align}
where $\ell=1/\sqrt{-\Lambda}$ is the AdS radius. The condition \eqref{no_ghost} conflicts \eqref{central_charge} under the limit $|\alpha|\to \infty$, i.e.~in NMG. Furthermore, the three conditions are inevitably inconsistent when $\sigma=-1$. On the other hand, all the conditions simultaneously hold when
\begin{align}
0<\alpha<1 \,,\quad
-2(1-\alpha) M_*^2<\Lambda<0
\end{align}
for $\sigma=+1$ and $\Lambda<0$. These inequalities imply that we need to choose the minus branch
\begin{align}
\Lambda&=-2M_*^2(1- \sqrt{1+\lambda} )
\,.
\end{align}
As a result, we obtain the conditions
\begin{align}
\sigma=+1\,, \quad
0<\alpha<1 \,, \quad -1+\alpha^2<\lambda<0
\,,
\end{align}
to preserve unitarity in both the bulk and the boundary.
A finite value of $\alpha$ implies that the torsion has a finite value and the higher curvature theory \eqref{TNMG} has the additional term $T\wedge K$. The torsion resolves the conflict of unitarity.

The inclusion of the torsion is originally motivated by localising the Poincar\'{e} group and the Poincar\'{e} group is obtained as a contraction of the AdS group. The vielbein and the spin connection can be combined to form a single connection for the AdS group.
PGT with a negative cosmological constant might be obtained as a low energy EFT of an AdS gauge theory by spontaneously breaking down the AdS group to the Lorentz group~\cite{MacDowell:1977jt,Stelle:1979aj}.\footnote{The general action of PGT has the local Lorentz symmetry but does not have either the local (A)dS symmetry or the local Poincar\'{e} symmetry. On the other hand, there is a case that the symmetry is enlarged from the Lorenz group to the (A)dS group or the Poincar\'{e} group in particular dimensions. The three-dimensional Einstein-Hilbert action with a negative cosmological constant in the vielbein formalism has the $SO(2,2)$ invariance and is equivalent to the Chern-Simons gauge theory for the AdS group~\cite{Achucarro:1987vz,Witten:1988hc}. In this case, gravity is indeed an AdS gauge theory and the symmetry is preserved. }  Intriguingly, the unitarity problem of the AdS/CFT correspondence in NMG is simply solved by adding the torsion which is expected to appear when the AdS group, the symmetry of the bulk, is localised and is spontaneously broken.

So far, we have started with the higher curvature theory~\eqref{TNMG} and found that \eqref{TNMG} can be recast in the form of a bigravity theory. Instead, one can assume the full bigravity (the zwei-dreibein gravity) action and integrate out the $f$-dreibein and the $g$-spin connection to obtain a more general theory of ghost-free higher curvature gravity. In general, the equation of motion of $\xi$ becomes non-linear. The resultant higher curvature theory must have an infinite number of higher curvature terms but these terms are specified by a finite number of coupling constants of the ghost-free bigravity theory.

\section{Four dimensions}
\label{sec_4d}
\subsection{Quadratic gravity}
In the NMG limit, one of the Einstein-Hilbert term in \eqref{bigravity_3d} must have the ``wrong'' sign, meaning that there exists a ghostly massless (or massive) spin-2 field. On the other hand, the existence of the torsion enables us to find an equivalent bigravity theory with the ``correct'' sign of the Einstein-Hilbert terms in three dimensions. This suggests the existence of a ghost-free higher curvature theory with the massless and massive spin-2 fields even in $D> 3$ dimensions thanks to the torsion. 

Let us consider a simple extension of TNMG in four dimensions,
\begin{align}
S=\frac{\Mpl^2}{4}  \intb \epsilon_{abcd} \Big[ &e^a \wedge e^b \wedge R^{cd} - \alpha e^a\wedge  T^b \wedge K^{cd} 
\nn
&- \frac{1}{M_*^2} e^a \wedge S^b \wedge R^{cd} \Big]
,
\label{TNMG_4d}
\end{align}
which only has the massive spin-2 mode in addition to the massless graviton at the level of linear perturbations just like the previous Lagrangian \eqref{TNMG} ~\cite{Sezgin:1979zf,Sezgin:1981xs}. We do not introduce the parameter $\sigma=\pm 1$ in four dimensions because the massless spin-2 field is dynamical. In the following we shall discuss non-linear properties of \eqref{TNMG_4d} by rewriting the action in the form of bigravity.

Following the same step as before, we find the equivalent action in the bigravity form,
\begin{align}
S_{\rm eq}=S_{\rm dRGT}[e,\Omega,f,\omega]+S_{\rm der}[e,\Omega,f,\omega]
\,,
\end{align}
where
\begin{align}
S_{\rm dRGT}&=\frac{\Mpl^2}{4}  \intb \epsilon_{abcd} 
\nn
&\times \big[ \alpha e^a \wedge e^b \wedge F^{cd}+(1-\alpha) f^a \wedge f^b \wedge R^{cd}
\nn
&\quad +\alpha (1-\alpha) m^2 e^a \wedge e^b \wedge (f^c-e^c) \wedge (f^d-e^d) \big],
\\
S_{\rm der}&=  \frac{(\alpha-1)\Mpl^2}{4} \intb \epsilon_{abcd} (f^a-e^a)\wedge (f^b-e^b) \wedge R^{cd}
,
\end{align}
with
\begin{align}
m^2=\frac{1-\alpha}{\alpha}M_*^2
\,.
\end{align}
As expected, we find the ghost-free bigravity action $S_{\rm dRGT}$ but we also have a derivative coupling $S_{\rm der}$. The essential difference between the three-dimensional case and the four-dimensional case is that the Einstein-Hilbert is non-linear in the vierbein in four dimensions. In three dimensions, we find the coupling $\xi \wedge R=(f-e)\wedge R $ in \eqref{eqTNMG} which generates the Einstein-Hilbert action of the $f$-variable, $f\wedge R$. In four (and higher than four) dimensions, after replacing the term $e\wedge S \wedge R$, we have the coupling $e \wedge \xi \wedge R$ which is linear in $\xi^a$ and thus cannot yield the Einstein-Hilbert action $f\wedge f\wedge R$. We need the non-linear term $\xi \wedge \xi \wedge R$ to find the Einstein-Hilbert action of the $f$-variable. As a result, the straightforward extension of TNMG in four dimensions (and $D>4$ as well) is not equivalent to the ghost-free bigravity theory; the resultant theory possesses the derivative interactions of the form $\xi \wedge \xi \wedge R = (f-e)\wedge (f-e) \wedge R$.

The derivative interaction $S_{\rm der}$ contributes to the dynamics from the cubic order of perturbations about the flat background, $e^a_{\mu}=f^a_{\mu}=\delta^a_{\mu},~\omega^{ab}{}_{\mu}=\Omega^{ab}{}_{\mu}=0$. In the free theory, the theory is free from the ghost when $0<\alpha<1$, and describes a massless spin-2 field and a massive spin-2 field. The question is whether or not $S_{\rm der}$ yields any pathology at non-linear orders. The coupling $(f-e)\wedge (f-e) \wedge R$ was already discussed by the paper \cite{deRham:2015rxa} in the context of massive gravity and it was shown that a new spin-1 dof appears though the well-known BD ghost~\cite{Boulware:1973my} does not appear. This additional mode must be a ghost or is at best strongly coupled in the flat spacetime limit, meaning that there is a cutoff of the theory which will be identified with $\Lambda_4=(\Mpl m^3)^{1/4}$ in Appendix \ref{sec_cutoff}. New physic is required at $\Lambda_4$ and the theory \eqref{TNMG_4d} is meaningful in energy scales only below the cutoff $\Lambda_4$.

The appearance of addition dof(s) at non-linear orders was also pointed out by Yo and Nester~\cite{Yo:2001sy} in the context of PGT although they do not present explicitly the analysis of PGT with the massive spin-$2^+$ mode. We thus call the additional ghostly modes of PGTs Yo-Nester (YN) ghosts. Similarly to the BD ghost problem of general massive gravities, general PGTs are suffered from the non-linear YN ghost(s).

\subsection{Ghost-free theory from bigravity}

Instead of introducing new physics at $\Lambda_4$, one may add appropriate counterterms to push the cutoff scale of \eqref{TNMG_4d} into a more higher scale.
Since $S_{\rm dRGT}$ is free from the ghost at fully non-linear orders, we can easily obtain the ghost-free theory by adding the counter term, $S_{\rm ct}=-S_{\rm der}$; that is,
\begin{align}
S_{\rm GF}=S_{\rm eq}+S_{\rm ct}=S_{\rm dRGT}
\,.\label{ghost-free_theory}
\end{align}
The lowest scale of interactions is then raised into $\Lambda_3=(\Mpl m^2)^{1/3}>\Lambda_4$ where $M_* \simeq m \ll \Mpl$ has been implicitly assumed in which the connection is weakly coupled. The cutoff of the theory is $\Lambda_3$ or can be larger than $\Lambda_3$ since no additional dof appears at non-linear orders. The dRGT theory can resolve the YN ghost problem of PGT as well as the BD ghost problem of massive gravity.

By the use of the auxiliary variables $\xi^a$ and $\chi^{ab}$, the bigravity action is written as
\begin{widetext}
\begin{align}
S_{\rm GF}[e,\omega,\xi,\chi]=\frac{\Mpl^2}{4} \intb \epsilon_{abcd} \Bigg[ 
& e^a \wedge e^b \wedge R^{cd}+2 \alpha \left(e^a \wedge T^b \wedge \chi^{cd}+\frac{1}{2}e^a \wedge e^b \wedge \chi^c{}_e \wedge \chi^{ed} \right)
\nn
&+(1-\alpha)\left( 2\xi^a \wedge e^b \wedge R^{cd}+\xi^a \wedge (\alpha m^2 e^b\wedge e^c+R^{bc})\wedge  \xi^d \right)\Biggl]
.
\end{align}
\end{widetext}
We have a non-linear term $\xi \wedge R \wedge \xi $ coming from the counter term. The equation of motion of $\chi^{ab}$ is still given by the same form
\begin{align}
\chi^{ab}=-K^{ab}
\,,
\end{align}
whereas the equation of motion of $\xi^a$ is
\begin{align}
\epsilon_{abcd} \left[ \xi^b \wedge ( \alpha m^2 e^c \wedge e^d+ R^{cd}) + e^b \wedge R^{cd} \right]=0
\,.
\label{eom_xi}
\end{align}
The solution of $\xi^a$ is schematically given by
\begin{align}
\xi(R) =-  \frac{R}{\alpha m^2} \left(1+ \frac{R}{\alpha m^2} \right)^{-1} 
\end{align}
and the explicit solution can be obtained perturbatively. Eliminating the auxiliary variables $\xi^a$ and $\chi^{ab}$, we find a ghost-free higher curvature theory,
\begin{align}
S_{\rm GF}[e,\omega]&=\frac{\Mpl^2}{4} \intb L_{\rm GF}
\,, \nn
L_{\rm GF}&=\epsilon_{abcd} \left( e^a\wedge e^b \wedge R^{cd}-\alpha e^a \wedge T^b \wedge K^{cd} \right)
\nn
& -\frac{1}{M_*^2} R \left(1+ \frac{R}{\alpha m^2} \right)^{-1} R 
\,,
\label{GF_high}
\end{align}
where 
\begin{align}
& R \left(1+ \frac{R}{\alpha m^2} \right)^{-1} R
=- \alpha m^2 \epsilon_{abcd}\xi^a(R) \wedge e^b \wedge R^{cd}
\nn
&= \epsilon_{abcd}  e^a \wedge S^b \wedge R^{cd} - \frac{1}{\alpha m^2} \epsilon_{abcd} S^a \wedge S^b \wedge R^{cd}
\nn
& +\mathcal{O}(R^4/(\alpha m^2)^2)
\,. 
\end{align}
There exists a non-linearly ghost-free higher curvature theory describing a massless spin-2 field and a massive spin-2 field if we add an infinite number of the appropriate higher curvature terms. The higher curvature terms can be given by a schematic form $R \left(1+ \frac{R}{\alpha m^2} \right)^{-1} R$.\footnote{We need to suppose the existence of the inverse $\left(1+ R/(\alpha m^2) \right)^{-1}$. As far as $|R|\ll \alpha m^2$, we can expand this expression in terms of $R/(\alpha m^2)$ and then we can safely conclude the equivalence between \eqref{GF_high} and the ghost-free bigravity. However, at $|R| \sim \alpha m^2$, the inverse can be singular, meaning that \eqref{GF_high} with $|R|\ll \alpha m^2$ and that with $|R|\gg \alpha m^2$ can be disconnected.}
The stability conditions simply lead to
\begin{align}
0<\alpha <1 \,,
\end{align}
as well as $\Mpl^2,M_*^2>0$.

It would be interesting that the higher curvature terms are suppressed by $\alpha m^2$ rather than $m^2$. When we take the limit $|\alpha|\to \infty$ with $m^2$ kept finite (note that $m^2 \to -M_*^2$ as $|\alpha| \to \infty$), we can ignore cubic and higher than cubic order terms, and find
\begin{align}
S_{\rm GF}[e] &= \frac{\Mpl^2}{4} \intb L_{\rm QG} +\mathcal{O}(\alpha^{-1}m^{-4})
\,, \nn
L_{\rm QG}&=\epsilon_{abcd} \left[ e^a\wedge e^b \wedge R^{cd}(e)-\frac{1}{M_*^2} e^a \wedge S^b(e) \wedge R^{cd}(e) \right]
\nn
&=\epsilon_{abcd} \left[ e^a\wedge e^b \wedge R^{cd}(e)+\frac{1}{m^2} e^a \wedge S^b(e) \wedge R^{cd}(e) \right]
, \label{QG_metric}
\end{align}
where the spin connection is integrated out by the use of $T^a=\mathcal{O}(\alpha^{-1}m^{-2})$. The last equality holds only under the $|\alpha|\to \infty$ limit. In the coordinate expression, the action is given by
\begin{align}
S_{\rm GF}[g] &= \frac{\Mpl^2}{2} \intb d^4x \sqrt{-g} \mathcal{L}_{\rm QG} + \mathcal{O}(\alpha^{-1} m^{-4})
\,, \nn
\mathcal{L}_{\rm QG}
&=R(g)+\frac{1}{M_*^2}\left[R_{\mu\nu}(g)R^{\mu\nu}(g) - \frac{1}{3} R(g)^2\right]
\nn
&=R(g)+\frac{1}{2M_*^2}\left[C_{\mu\nu\rho\sigma}(g)C^{\mu\nu\rho\sigma}(g)-R^2_{\rm GB}(g) \right]
, \label{QG_metric_component}
\end{align}
where $C_{\mu\nu\rho\sigma}(g)$ is the Weyl tensor and $R^2_{\rm GB}(g)$ is the Gauss-Bonnet term.
We obtain a quadratic gravity in four dimensions. Since this action is obtained as a scaling limit, the dynamical degrees of freedom are still the massless spin-2 field and the massive spin-2 field only. However, in the limit $|\alpha|\to \infty$, either one of the ``Planck masses'', $M_g^2=\alpha \Mpl^2$ or $M_f^2=(1-\alpha)\Mpl^2$ must be negative. Either the massless spin-2 or the massive spin-2 is a ghost.

Even for a finite $\alpha$, one can obtain the same action \eqref{QG_metric_component} as an effective description when one is interested in energy scales below $m$. By integrating out the torsion, in low energies $E\ll m$, we find
\begin{align}
S_{\rm QG}&=\frac{\Mpl^2}{2}\intb d^4x \sqrt{-g} \mathcal{L}_{\rm QG} + \mathcal{O}(m^{-4})
\,.
\label{QG_low_E}
\end{align}
Note, however, that two actions \eqref{QG_metric_component} and \eqref{QG_low_E} are physically different. In \eqref{QG_metric_component}, we just take the limit $|\alpha|\to \infty$ and do not assume the energy scale. On the other hand, the action \eqref{QG_low_E} is obtained as an effective description of the theory \eqref{GF_high} in low energies $E\ll m$. Although both \eqref{QG_metric_component} and \eqref{QG_low_E} contain the massive spin-2 ghost because of the higher curvature term, the ghost in \eqref{QG_metric_component} is a ``real'' ghost whereas the ghost in \eqref{QG_low_E} appears as a result of breakdown of the approximation. The $\mathcal{O}(m^{-4})$ terms cannot be ignored at $E\sim m$.

We can consider low energy scattering problems based on the effective action \eqref{QG_low_E}. It is recognized that not all apparently consistent low energy EFTs admit a ``standard'' UV completion~\cite{Adams:2006sv}. The standard axioms of the scattering amplitude, namely unitarity, Lorentz invariance, causality and locality, lead to non-trivial constraints on non-gravitational EFTs called the positivity bounds~\cite{Adams:2006sv}. As for gravitational EFTs, the positivity bounds are derived when an expected UV behaviour of the amplitude, the Regge behaviour, is assumed~\cite{Tokuda:2020mlf} (see also \cite{Hamada:2018dde,Bellazzini:2019xts,Alberte:2020jsk}). A detailed study~\cite{deRham:2019ctd} concludes the positive sign of the Weyl square term, $M_*^2>0$, under the assumption that the graviton $t$-channel pole can be ignored, and this assumption is justified when $M_*$ is much lower than the scale of UV completion of gravity~\cite{Tokuda:2020mlf}. The requirement of the positivity is consistent with the requirement of the stability condition of the ghost-free theory~\eqref{ghost-free_theory}, as it should be.

As mentioned in the three-dimension case, we can assume the full four-dimensional bigravity theory and then obtain the equivalent higher curvature theory by integrating out the vierbein and the spin connection. A similar discussion was already made in~\cite{Paulos:2012xe}, considering the general $D$ dimensional bigravity theory; however, unitarity needs to be explicitly broken to find a corresponding higher curvature theory in their analysis although the BD ghost remains absent. This is because they work in the metric formalism where the torsion vanishes, which corresponds to the limit $|\alpha|\to \infty$ of our case. When the torsion is allowed and $0<\alpha<1$ is satisfied, unitarity can be preserved, on the other hand. This conclusion is expected to hold in generic $D> 4$ dimensions.

\section{Inclusion of additional dof}
\label{sec_add_scalar}
Since generic PGTs contain not only the massive spin-$2^+$ mode but also other spin modes, it is natural to consider extensions of the theory \eqref{GF_high} to theories including additional massive dofs. In this section, we present an example of the extension that contains an additional spin-0 mode in four dimensions.

The dynamical spin-$0^-$ mode of the spin connection shows up around the flat background when the term $d^4x |e| \Xcal^2$ is added where $\Xcal$ is so-called the Holst scalar defined by
\begin{align}
\Xcal:=-\frac{1}{2}\epsilon^{\mu\nu\rho\sigma}R_{\mu\nu\rho\sigma}
\,.
\end{align}
However, the naive inclusion of the $\Xcal^2$ term should break the special structure of the ghost-free theory. The YN ghost(s) must reappear, in general.

We study the action,
\begin{align}
&S_{2^+ \& 0^-}[e,\Omega,f,\omega,\varphi]
\nn
&=\frac{\Mpl^2}{4}\intb \epsilon_{abcd}(\alpha e^a\wedge e^b \wedge F^{cd}+(1-\alpha) f^a \wedge f^b \wedge R^{cd})
\nn 
&+\frac{\Mpl^2}{2}\intb \varphi f^a \wedge f^b \wedge R_{ab} +S_{\rm mass}[e,f,\varphi]\,,
\label{2+0-}
\end{align}
where $\varphi$ is a parity odd scalar field. $S_{\rm mass}$ is an extension of the dRGT mass terms,
\begin{align}
S_{\rm mass}=\frac{\Mpl^2M_*^2}{4} \sum_{A,B,C,D} \intb  c_i(\varphi) \epsilon_{abcd} e^a_A \wedge e^b_B \wedge e^c_C \wedge e^d_D
\,,
\end{align}
where $e^a_A=\{e^a, f^a\}$ and $c_i(\varphi)$ are arbitrary functions of $\varphi$. The spin connections $\Omega^{ab}$ and $\omega^{ab}$ in \eqref{2+0-} are non-dynamical and thus can be integrated out. We then obtain
\begin{align}
&S_{2^+\&0^-}[e,f,\varphi]
\nn
&=\frac{\Mpl^2}{4}\intb \epsilon_{abcd}(\alpha e^a\wedge e^b \wedge R^{cd}(e)
\nn
&\qquad \qquad \qquad \quad +(1-\alpha) f^a \wedge f^b \wedge R^{cd}(f) )
\nn 
&-\frac{3\Mpl^2(1-\alpha)}{4}\intb d^4x |f^a_{\mu}| \frac{(\partial \varphi)_f^2}{1+\varphi^2} +S_{\rm mass}[e,f,\varphi]\,,
\label{2+0-a}
\end{align}
where $(\partial \varphi)_f^2=\eta^{ab}f_a^{\mu}f_b^{\nu}\partial_{\mu}\varphi \partial_{\nu}\varphi=f^{\mu\nu}\partial_{\mu}\varphi \partial_{\nu}\varphi$. When we introduce the canonically normalised field $\theta$ via
\begin{align}
\varphi=\sinh \theta
\,,
\end{align}
the action takes
\begin{align}
&S_{2^+\&0^-}[e,f,\theta]
\nn
&=\frac{\Mpl^2}{4}\intb \epsilon_{abcd}(\alpha e^a\wedge e^b \wedge R^{cd}(e)
\nn
&\qquad \qquad \qquad \quad +(1-\alpha) f^a \wedge f^b \wedge R^{cd}(f) )
\nn 
&-\frac{3\Mpl^2(1-\alpha)}{4}\intb d^4x |f^a_{\mu}| (\partial \theta)_f^2+S_{\rm mass}[e,f,\sinh \theta]\,.
\label{2+0-b}
\end{align}
in which the scalar field $\theta$ has the canonical kinetic term coupling to the $f$-variables and has interactions with both $g$- and $f$-variables through the dRGT mass terms. It is known that a matter field coupling to both $g$- and $f$-variables reintroduces the BD ghost, in general~\cite{Yamashita:2014fga,deRham:2014naa}. However, the interactions of \eqref{2+0-b} is a special one that is free from the BD ghost. The similar Lagrangian is introduced by~\cite{DeFelice:2017oym} as the chameleon bigravity because the mass of the spin-2 field depends on the scalar field $\theta$ through the functions $c_i(\varphi)$. The action \eqref{2+0-} is recast in the form of the chameleon bigravity and thus is free from the ghost at non-linear orders.

We then back to the Lagrangian \eqref{2+0-} and introduce the auxiliary variables $\xi^a$ and $\chi^{ab}$. Since $S_{\rm mass}[e,f,\varphi]$ has no derivatives of the fields, $S_{\rm mass}[e,\xi,\varphi]$ is algebraic in $\xi^a$ and $\varphi$. The new term $\varphi f\wedge f \wedge R$ becomes
\begin{align}
\varphi f^a \wedge f^a \wedge R_{ab}=\varphi (e^a+\xi^a)\wedge (e^b+\xi^b)\wedge R_{ab}
\,,
\end{align}
and is also algebraic in $\xi^a$ and $\varphi$. The Einstein-Hilbert actions are the same as before. Therefore, the variables $\chi^{ab},\xi^a$, and $\varphi$ are non-dynamical variables and can be integrated out. The equations of motion of $\chi^{ab}$ takes exactly the same form $\chi^{ab}=-K^{ab}$, leading to the coupling $e\wedge K \wedge T$. We can also eliminate the non-dynamical variables $\xi^a$ and $\varphi$ by solving their equations of motion at least perturbatively. As an example, we suppose the interaction,
\begin{align}
S_{\rm mass}=\frac{\Mpl^2M_*^2}{4}
\int \epsilon_{abcd}\Big[&-\frac{\varphi^2}{8\alpha_{0^-} } e^a \wedge e^b \wedge e^c \wedge e^d 
\nn
&+ (1-\alpha)^2e^a \wedge e^b \wedge \xi^c \wedge \xi^d \Big]
,
\end{align}
where $\alpha_{0^-}$ is a dimensionless constant. The mass of the spin-$0^-$ mode is $m_{0^-}^2=(1-\alpha)M_*^2/\alpha_{0^-} $ about the flat background.
Up to leading order, the action is given by
\begin{align}
S_{2^+\&0^-}[e,\omega]&=
\frac{\Mpl^2}{4}\intb \epsilon_{abcd} \Big[ e^a \wedge e^b \wedge R^{cd}
\nn
&-\alpha e^a \wedge T^b \wedge K^{cd} -\frac{1}{M_*^2}e^a \wedge S^b \wedge R^{cd}  \Big]
\nn
& +\frac{\alpha_{0^-} \Mpl^2}{12M_*^2}\intb d^4x |e| \Xcal^2 
 +\mathcal{O}(M_*^{-4})
 \,.
\end{align}
We obtain a ghost-free higher curvature theory with the massive spin-$2^+$ and spin-$0^-$ modes.

One may consider the coupling $\varphi e\wedge e\wedge F$ instead of $\varphi f \wedge f \wedge R$. In the bigravity frame, the only difference from \eqref{2+0-a} is that the kinetic term of $\varphi$ appears in the $g$-sector. However, one should recall the relation \eqref{dif_cur} (the same relation holds in four dimensions). Hence, there is the coupling $\varphi e^a \wedge e^b \wedge D\chi_{ab}$ when the variables $\chi^{ab},\xi^a,\varphi$ are used, and then the action is no longer algebraic in either $\varphi$ or $\chi^{ab}$. The variables $\xi^a,\chi^{ab}$ and $\varphi$ may be integrated out formally but the resultant theory must be non-local. Although this non-local theory has the spin-0 dof, this dof should be regarded as the dof of $\varphi$ rather than a part of the spin connection. To obtain the local higher curvature gravity, the kinetic term of $\varphi$ should exist in the $f$-sector.

\section{Summary and discussions}
\label{summary}
In the present paper, we have shown the equivalence between a particular class of PGT and the ghost-free bigravity in a vacuum. Although the similar relation was shown in \cite{Paulos:2012xe} by taking a scaling limit, the limit necessarily breaks unitarity; either the massless spin-2 mode or the massive spin-2 mode is a ghost. On the other hand, we have shown that even if both of the massless and massive spin-2 modes are not ghostly modes, the equivalence holds thanks to the torsion. The limit considered in \cite{Paulos:2012xe} is the limit to vanish the torsion. We have also presented an example extension that includes an additional massive spin-0 degree of freedom in four dimensions. The ghost-free bigravity and its extension can arise from higher curvature corrections to the Einstein-Hilbert action. The higher dimensional scenarios predict multiple massive spin-2 fields. Hence, if the number of the massive spin-2 states is clarified, we may argue whether the massive spin-2 is originated from the higher dimensions or the higher curvatures. From the point of view of PGT, the equivalence provides counterexamples to the previous argument about ghost-free PGTs. 

Let us comment on matter couplings. In the context of PGT, spin currents, as well as energy-momentum currents, are sources for gravity. The minimal couplings between the spin connection and matter fields are introduced by replacing the partial derivative $d$ with the covariant derivative $D$. Regarding the bosonic fields that have no Lorentz indices such as scalar, vector, and form fields, the covariant exterior derivative is the same as the exterior derivative, $DA^p=dA^p$ where $A^p$ is a $p$-form, meaning that there is no coupling between the spin connection and these matter fields. On the other hand, fermionic fields $\psi$ and tensor fields of the Lorentz group $T^{ab\cdots}$ have the coupling through the covariant derivatives, $D\psi, DT^{ab\cdots}$. Spin currents for these matter fields become sources of the spin connection. If one adopts the minimal couplings, the equivalent bigravity theory has apparently peculiar couplings: the matter fields are coupled with the $g$-vielbein and the $f$-spin connection. Energy-momentum currents are sources for the $g$-variable while spin currents are sources for the $f$-variable. A new type of doubly matter coupling is expected.

The connection between massive gravity and PGT could serve interesting implications not only for studies about massive gravity and PGT but also for the AdS/CFT correspondence. The massive spin-2 mode can play the role of dark matter~\cite{Aoki:2016zgp,Babichev:2016hir,Babichev:2016bxi,Aoki:2017cnz,Aoki:2017ffl,Marzola:2017lbt} and we can rediscuss these scenarios based on PGTs in which the new type of matter coupling is predicted.
It would be also intriguing to see implications of higher curvature terms for the AdS/CFT correspondence. In three dimensions, the torsionless quadratic gravity, known as NMG, has a conflict between unitarity in the bulk and the dual CFT. This conflict is simply resolved by adding the torsion. Furthermore, higher dimensional extensions are straightforward. Supposing the ghost-free bigravity theory in generic $D$-dimensions in the vielbein formalism, one may obtain the corresponding higher curvature theory by integrating out the $g$-spin connection and the $f$-vielbein. One can discuss holographic implications of higher curvature terms in generic dimensions without suffering from the ghost.

In four-dimensional bigravity, the Schwarzschild(-AdS) black hole (BH) is a vacuum solution and Schwarzschild BH becomes unstable when the horizon radius is smaller than the Compton wavelength of the massive spin-2 field~\cite{Babichev:2013una,Brito:2013wya}. At the same time, hairy BH solutions have been found~\cite{Volkov:2012wp,Brito:2013xaa,Gervalle:2020mfr} and these BHs are natural candidates for the final state of the instability of the Schwarzschild BH. Since the ghost-free bigravity is shown to be equivalent to the ghost-free higher curvature gravity, the same phenomena should occur: a small BH may exhibit a phase transition and the threshold is determined by the scale that the higher curvature terms are important. On the other hand, in three dimensions, the situation could be different since the Banados-Teitelboim-Zanelli BH~\cite{Banados:1992wn} is locally AdS and then would be stable even in the presence of the massive spin-2 field, suggesting no phase transition. It must be interesting to investigate (non-)existence of hairy BH solutions in general dimensions and show whether such a phase transition is indeed present or absent in bigravity/higher curvature gravity which we leave for a future study.

\section*{Acknowledgments}
We would like to thank Shinji Mukohyama for useful discussions. We acknowledge the xTras package~\cite{Nutma:2013zea} which was used for confirming tensorial calculations. 
The work of K.A. was supported in part by Grants-in-Aid from the Scientific Research Fund of the Japan Society for the Promotion of Science, No.~19J00895 and No.~20K14468. 

\appendix

\section{Cutoff of the quadratic gravity}
\label{sec_cutoff}
As shown in the main text, an infinite number of the higher curvature terms can yield a ghost-free completion of the higher curvature gravity without introducing any additional dof. Instead, one may consider another scenario with the help of a heavy dof which changes the theory at an energy scale well-below the scale of the ghostly operator. Even if a theory contains a ghostly operator, the theory can be thought of as an EFT. 

The quadratic curvature terms are particularly interesting in four dimensions because these terms have an additional local symmetry, the scale invariance. Since the curvature 2-form is defined by the spin connection, $R^{ab}$ is invariant under the following scale transformation
\begin{align}
e^a \to e^{\Omega(x)} e^a\,, \quad \omega^{ab} \to \omega^{ab}
\,, \label{Weyl_trans}
\end{align}
where $\Omega(x)$ is an arbitrary function. The quadratic curvature terms such as $e\wedge S \wedge R$ are invariant under the transformation \eqref{Weyl_trans}. If these quadratic curvature terms become dominant at some energy scale, there exists an approximate scale invariance which might be used to cause the inflationary universe at such an energy scale~\cite{Aoki:2020zqm}, just like the Starobinsky inflation~\cite{Starobinsky:1980te}.\footnote{Precisely speaking, the successful inflationary scenario should require a spin-0 dof so the present model \eqref{TNMG_4d} itself may not be used for inflation.} We, therefore, rediscuss the quadratic model \eqref{TNMG_4d} and clarify the scale of the lowest ghostly operator.

We follow the decoupling limit analysis used in the paper \cite{deRham:2015rxa} (see also \cite{Ondo:2013wka,Fasiello:2013woa}).
As we have shown, the equivalent theory of the quadratic gravity \eqref{TNMG_4d} is $S_{\rm eq}=S_{\rm dRGT}+S_{\rm der}$. For later convenience, we use the Planck mass of the $f$-sector $M_f^2=(1-\alpha)\Mpl^2$ instead of the original Planck mass $\Mpl$. By means of three parameters $M_f,m$ and $\alpha$, the equivalent action is
\begin{align}
S_{\rm dRGT}&=\frac{M_f^2}{4}  \intb \epsilon_{abcd} \biggl[ \frac{\alpha}{1-\alpha} e^a \wedge e^b \wedge F^{cd}+ f^a \wedge f^b \wedge R^{cd}
\nn
&\qquad  \qquad \qquad 
+\alpha  m^2 e^a \wedge e^b \wedge \xi^c \wedge \xi^d \biggl],
\\
S_{\rm der}&= - \frac{M_f^2}{4} \int_{\Mcal_4} \epsilon_{abcd} \xi^a \wedge \xi^b \wedge R^{cd}
\,,
\end{align}
where $\xi^a=f^a-e^a$ is understood. We then define the $\Lambda_n$ decoupling limit by 
\begin{align}
M_f \to \infty \,, \quad m \to 0
\end{align}
while keeping
\begin{align}
\Lambda_n=(M_f m^{n-1})^{1/n} \to {\rm finite}
,
\end{align}
under which higher dimensional operators suppressed by $\Lambda_{n'}$ with $n'<n$ vanish.

In the bigravity action $S_{\rm eq}[e,\Omega,f,\omega]$, there are two sets of the vierbeins and the spin connections but there is only one set of the diffeomorphism invariance and the local Lorentz invariance. To ``recover'' two sets of the invariances, we shall introduce St\"uckelberg fields as
\begin{align}
e^a&=\Lambda^a{}_b e^b{}_c(\phi) d\phi^c
\,, \\ 
\Omega^{a}{}_{b}&=d\phi^e \left[ \Lambda^a{}_c \Omega^c{}_{de}(\phi)(\Lambda^{-1})^d{}_b+\Lambda^a{}_c \partial_e (\Lambda^{-1})^c{}_b \right]
\end{align}
where $\phi^a$ are four scalar fields and $\Lambda^a{}_b$ is the Lorentz St\"uckelberg field, respectively.
We then define the canonically normalised fields about the flat background via
\begin{align}
\omega^{ab}&=\frac{1}{M_f }\mu^{ab}
\,, \quad
f^a=\one^a+\frac{1}{M_f } h^a
\label{f_pert}
\,, \\
\Omega^{ab}{}_c &=\frac{1}{M_g } \nu^{ab}{}_c 
\,, \quad
e^a{}_b=\delta^a_b+\frac{1}{M_g }  \gamma^a{}_b
\,, 
\label{g_pert}
\\
\phi^a&=x^a+\frac{1}{M_f m} A^a +\frac{1}{M_f m^2}\partial^a \pi
\,, 
\label{phi_pert}
\\
\Lambda^a{}_b&=\exp\left[\frac{1}{M_f m} \lambda^a{}_b \right]
\,,
\label{lam_pert}
\end{align}
where $M_g^2=\alpha \Mpl^2=\frac{\alpha}{1-\alpha}M_f^2$ and $\one^a=\delta^a_{\mu}dx^{\mu}$. We now have seven variables 
\begin{align*}
\{ \mu^{ab}{}_{\mu}dx^{\mu}, h^a{}_{\mu}dx^{\mu}, \nu^{ab}{}_c, \gamma^{a}{}_b, A^a, \pi,\lambda^a{}_b\}
\end{align*}
and expand the action in terms of them.

One can perform the general decoupling limit analysis of the bigravity theory~\cite{Fasiello:2013woa}. However, for the sake of simplicity, we further consider the limit
\begin{align}
\alpha \to 1
\end{align}
which is the weak coupling limit of the $g$-fluctuations, namely $M_g^2 \gg M_f^2$. The fluctuations of the $g$-variables, $\nu^{ab}{}_c$ and $\gamma^a{}_b$, are then decoupled from those of the $f$-variables and the St\"uckelberg fields. Hence, the action for the $f$-variables and the St\"uckelberg fields under the limit $\alpha \rightarrow 1$ is simply obtained by substituting
\begin{align}
e^a= \Lambda^a{}_b d\phi^b\,, \quad F^{ab}=0\,,
\end{align}
and \eqref{f_pert}, \eqref{phi_pert}, and \eqref{lam_pert} into the action. Hereinafter we only consider the fluctuations of the $f$-variables and St\"uckelberg fields.

It is known that the dRGT mass term is free from the BD ghost, and the lowest strong coupling scale is $\Lambda_3$ around the Minkowski spacetime. Indeed, ignoring boundary terms, we obtain
\begin{align}
&M_f^2 \intb \epsilon_{abcd} f^a \wedge f^b \wedge R_{(f)}^{cd}
\nn&=\intb
\epsilon_{abcd}\left( 2 h^a \wedge d\mu^{bc}\wedge \one^d +  \mu^{a}{}_e \wedge \mu^{eb} \wedge \one^c \wedge \one^d +\cdots \right)
,\\
&M_f^2 m^2 \intb \epsilon_{abcd}e^a \wedge e^b \wedge \xi^c \wedge \xi^d
\nn
&=\intb
\epsilon_{abcd}\Big( -2h^a \wedge d\partial^b \pi \wedge \one^c \wedge \one^d
\nn
&\qquad \qquad \quad~
+2 dA^a \wedge \lambda^b{}_e \one^e \wedge \one^c \wedge \one^d
\nn 
&\qquad \qquad \quad~
+\lambda^a{}_e \one^e\wedge \lambda^b{}_f \one^f \wedge \one^c \wedge \one^d +\cdots
\Big)
\end{align}
where $\cdots $ are non-linear operators suppressed by $\Lambda_3$ or higher than $\Lambda_3$. The $\cdots$ are irrelevant when we consider the $\Lambda_n$ decoupling limit with $n>3$. On the other hand, the derivative coupling $\xi \wedge \xi\wedge R$ is given by
\begin{align}
 M_f^2 \intb \epsilon_{abcd}& \xi^a \wedge \xi^b \wedge R^{cd} 
 \nn
=\intb \epsilon_{abcd} \Big[& \frac{1}{\Lambda_5^5} d\partial^a \pi \wedge d \partial^b \pi \wedge d\mu^{cd}
\nn
&+\frac{2}{\Lambda_4^4} d\partial^a\pi \wedge \lambda^b{}_e \one^e \wedge d\mu^{cd}_{(f)} \Big]
 +\cdots\,,
\end{align}
where $\cdots$ are operators suppressed by higher than $\Lambda_4$. Thanks to the exterior products, the first term, the operator at $\Lambda_5$, possesses the total derivative structure. Although the naive cutoff scale is $\Lambda_5$ which is known as the lowest cutoff of general massive gravity, the cutoff is raised into $\Lambda_4$ because of the special structure of the interaction.

However, the operator at $\Lambda_4$, namely $d\partial \pi \wedge \lambda \wedge d\mu$, is a ghostly operator as shown by \cite{deRham:2015rxa}. After performing integration by parts, the variable $\mu^{ab}$ is a non-dynamical variable. We can thus eliminate $\mu^{ab}$ from the action. The existence of the operator  $d\partial \pi \wedge \lambda \wedge d\mu$ implies that the Lorentz St\"uckelberg field $\lambda^{a}{}_b$ gets a kinetic term of the form,
\begin{align}
\frac{1}{\Lambda_4^8}(\partial \lambda)^2 (\partial^2 \pi)^2
\,, \label{NL_kin}
\end{align}
after integrating out $\mu^{ab}$ where the indices are omitted for the notational simplicity. On the other hand, $\lambda^a{}_b$ does not have a quadratic kinetic term; thus, $\lambda^a{}_b$ must be a ghostly dof or at best strongly coupled. Note that in the paper \cite{deRham:2015rxa} they rescale the coupling constant of $\xi \wedge \xi \wedge R$ and then take the $\Lambda_3$ decoupling limit. In the present case, on the other hand, there is no free parameter to rescale the coupling constant of $\xi \wedge \xi \wedge R$. We thus find the ghostly mode at the scale $\Lambda_4$. 

There might be a possibility that the variable $\lambda^a{}_b$ represents a healthy spin-1 dof. Since the generic PGTs involve a spin-1 dof, the strong coupling issue of $\lambda^a{}_b$ can be resolved by adding appropriate quadratic torsion and quadratic curvature terms. Inclusion of additional terms, however, may change the structure of the theory and then the YN ghost(s) could appear, in general. It would be interesting to see whether the ghost appears in such models and whether the ghost can be eliminated by introducing appropriate counterterms as we did in the main text.

In summary, the cutoff of the quadratic gravity \eqref{TNMG_4d} is $\Lambda_4 \simeq (\Mpl m^3)^{1/4}$. As far as $m \ll \Mpl$, there is an interesting energy regime $m < E \ll \Lambda_4$ in which the quadratic curvature term may provide relevant effects and the theory could be still meaningful. Phenomenological implications of the spin-2 mode of the quadratic PGT are recently discussed in~\cite{Aoki:2020zqm} in the context of inflation and is studied in~\cite{Nikiforova:2016ngy,Nikiforova:2018pdk} as for the late time cosmology.

\bibliography{ref}
\bibliographystyle{JHEP}

\end{document}